\documentclass[a4paper,11pt]{article}
\pdfoutput=1
\usepackage{a4wide}
\usepackage{amsfonts}
\usepackage{amssymb}
\usepackage{amsmath}
\usepackage{graphicx}
\usepackage{multirow}
\usepackage{float}
\usepackage{booktabs}
\usepackage{array}
\usepackage{color}
\usepackage[small,bf]{caption}
\usepackage{subcaption}
\usepackage{cite}
\usepackage{mciteplus}
\usepackage[svgnames]{xcolor}

\usepackage{hyperref} 
\usepackage{appendix}
\usepackage{verbatim}
\usepackage{textcomp}
\usepackage{cleveref}
\usepackage{braket}
\usepackage{bbm}
\usepackage{slashed}
\usepackage{units}
\allowdisplaybreaks

\numberwithin{equation}{section}

\newcommand{\ag}{\alpha_g}
\newcommand{\atilde}{\tilde{a}}
\newcommand{\mpbh}{M_{\text{PBH}}}
\newcommand{\fpbh}{f_{\text{PBH}}}
\newcommand{\mphi}{m_\phi}
\newcommand{\mV}{m_V}
\newcommand{\gvp}{g_{\nu \phi}}
\newcommand{\gvV}{g_{\nu V}}

\DeclareMathOperator{\diff}{\text{d}}

\begin{document}
\begin{titlepage}

\begin{center}
{
\bf\LARGE Neutrino superradiance constraint on\\[0.2em]
asteroid-mass PBH Dark Matter and beyond
}
\\[8mm]
Yi-Xuan Lin$^{\, a}$, \footnote{E-mail: \texttt{lindalin@gapp.nthu.edu.tw}}
Priyanka Sarmah$^{\, a,\, b, \, c}$ \footnote{E-mail: \texttt{priyankasarmah@gauhati.ac.in}}
and 
Martin Spinrath$^{\, a}$ \footnote{E-mail: \texttt{spinrath@phys.nthu.edu.tw}}
\\[1mm]
\end{center}

\vspace*{0.50cm}
{\centering \it
$^{a}$ Department of Physics, National Tsing Hua University,\\ 
    Hsinchu, 30013, Taiwan\\[0.2cm]
$^{b}$ Center for Theory and Computation,
	National Tsing Hua University,\\ 
    Hsinchu, 30013, Taiwan\\[0.2cm]
$^{c}$ Department of Physics, Gauhati University,\\
    Assam, 781014, India.\\
}
\vspace*{1.20cm}

\begin{abstract}
\noindent
Primordial Black Holes (PBHs) are an attractive candidate for
Dark Matter (DM) and there has been extensive experimental efforts
to look for them. The asteroid-mass window,
$\mpbh \sim 10^{17} - 10^{23}$~g, is particularly interesting, since PBHs in 
this range may still constitute all of DM. In this work, we study a scenario in which
rotating PBHs are surrounded by superradiantly produced boson clouds that emit an
approximately steady and nearly monochromatic flux of neutrinos in the few MeV range.
We compute both the Galactic and extragalactic neutrino fluxes from such PBH populations
and compare them with existing low-energy antineutrino limits from Borexino, KamLAND,
and Super-Kamiokande. For scalar bosons, we find that these neutrino searches can
strongly constrain a significant part of the asteroid-mass window and extend to somewhat
larger masses. For instance, for rapidly rotating PBHs with spin $\tilde{a}=0.9$ and gravitational
fine-structure coupling $\ag = 0.25$, the strongest bound on dark matter fraction
reaches approximately $\fpbh \sim 10^{-7}$ around $\mpbh \sim 2 \times 10^{22}$~g for a Yukawa coupling $\gvp = 10^{-4}$.
These constraints can be significantly stronger than existing microlensing limits
in the same mass range. Our results provide a complementary neutrino probe of PBH DM,
distinct from previous neutrino constraints based mainly on Hawking evaporation
from lighter PBHs.

\end{abstract}

\end{titlepage}
\setcounter{footnote}{0}

\section{Introduction}

Primordial black holes (PBHs) are considered one of the most economical and well-motivated possibilities
for dark matter (DM). They may have formed in the early universe from the collapse of large primordial 
overdensities or beyond standard cosmological mechanisms 
\cite{Carr:1974nx,Sasaki:2018dmp, Hawking:1982ga,Kodama:1982sf, Baker:2021nyl,Gross:2021qgx}.
Depending on their formation history, PBHs can span a broad range of masses. This catches attention as the
PBHs  may  constitute either a subdominant component or the entirety of the DM abundance,
for recent reviews see, for instance, \cite{pbh1Carr:2020gox, pbh2Carr:2020xqk, pbh3Carr:2021bzv, pbh4Green:2020jor}. After the first observation of a black hole merger by LIGO \cite{LIGOScientific:2016aoc}
and subsequent observations PBHs obtained a renewed interest \cite{ligopbh1Bird:2016dcv, ligopbh2Clesse:2016vqa}.

Among the possible mass ranges of PBHs, the so-called asteroid-mass PBHs draw particular attention because
the mass range $\mpbh \sim 10^{17}$ -- $10^{23}$~g  remains one of the least constrained windows in which 
PBHs may constitute all of DM, see, for instance, \cite{asteroid1Tinyakov:2024mcy}.
PBHs below $\sim 10^{17}$~g are mainly constrained by Hawking evaporation, while various microlensing 
searches provide strong bounds on masses above $\sim 10^{23}$~g; the intermediate window, referring to the 
asteroid mass range, is difficult to probe because such PBHs are too light for standard lensing and too heavy to 
evaporate efficiently today. This makes the asteroid-mass window a particularly interesting target for current and
future indirect and multi-messenger searches.

Various studies proposed the possibility of photon and multi-messenger searches, primarily originating 
from Hawking evaporation of PBHs, to probe parts of the asteroid-mass region PBH window. In particular, 
Refs.~\cite{Ray:2021mxu, Tan:2024nbx} show that lower parts of the asteroid-mass PBH window
$\mpbh \sim 10^{16}$ -- $10^{18}$~g,
can be tested through high-energy particles produced by Hawking evaporation.
While Ref.~\cite{Tan:2024nbx} reports a exclusion of $\fpbh = 1$ around 
$\mpbh \simeq 2.5 \times 10^{16}$ -- $3 \times 10^{18}$~g
using the existing cosmic X-ray background, Ref.~\cite{Ray:2021mxu} shows future MeV gamma-ray telescopes such
as AMEGO can probe Hawking photons from PBHs around  $\mpbh \sim 10^{17}$~g, and exclude non-spinning monochromatic 
PBHs as all of the DM up to $\mpbh \simeq 7 \times 10^{17}$~g at $95\%$~CL.

Neutrino searches have also been used to constrain light evaporating PBHs through the Hawking-emitted
neutrino flux see, e.g., \cite{nupbh1Dasgupta:2019cae, nupbh2Bernal:2022swt, nupbh3Lunardini:2019zob, nupbh4Singh:2026hnk, nupbh5Mukhopadhyay:2026lmz}.
These studies have considered both low- and high-energy neutrino signals in the context of existing and 
future neutrino detectors, as well as their cosmological impacts, such as contributions to N$_{\text{eff}}$ 
and heating of the intergalactic medium. In contrast, our work focuses on neutrinos generated
via the superradiance mechanism which could be detected on Earth.

The authors of
Ref.~\cite{Chen:2023vkq} showed that rotating black holes surrounded by superradiant boson clouds can behave as
efficient ``fermion factories'', neutrino factories in particular.  Black-hole superradiance occurs
when a light bosonic field forms a bound state 
around a rotating black hole and extracts its angular momentum, leading to the growth of a bosonic cloud
\cite{1971JETPL..14..180Z, 1971NPhS..229..177P, 1972JETP...35.1085Z}.
If this cloud couples to fermions, neutrinos can be produced through parametric production or Schwinger pair 
production, depending on whether the cloud is scalar or vector in nature. The energy carried away by
the produced fermions can also quench the superradiant growth, leading to a saturated phase in which
the energy extracted from the black hole is balanced by the outgoing fermion flux. This gives a steady neutrino 
signal whose strength depends on the boson--neutrino coupling and the black-hole parameters. Following 
Ref.~\cite{Chen:2023vkq}, the consecutive works \cite{Banerjee:2024knt, Banerjee:2025ddr} have used
this mechanism to constrain the parameter space of light/ultra-light bosons using neutrino and gravitational-wave
signals.

Besides the works \cite{Chen:2023vkq,Banerjee:2024knt,Banerjee:2025ddr}
superradiance in the context of PBHs has also been explored before as a probe of light
bosons and PBH DM in \cite{srpbh1Ferraz:2020zgi,srpbh2Branco:2023frw,srpbh3Dent:2024yje}.
In particular, Ref.~\cite{srpbh3Dent:2024yje} combined
superradiance-induced axion-like-particle line signals with Hawking
radiation and microlensing, showing that SXI, JWST, and AMEGO-X can test
complementary regions of the asteroid mass PBH $(\mpbh,\fpbh)$ and 
axion-like particle parameter spaces.
While Ref.~\cite{nupbh5Mukhopadhyay:2026lmz} probed asteroid-mass PBHs
using high-energy neutrinos coming out of Hawking radiation, our work focuses on the MeV neutrino flux
generated by superradiant boson clouds around asteroid-mass PBHs. We compute
the contribution from both Galactic and extragalactic PBH populations and
show that the predicted signal can exceed the existing low-energy neutrino
bounds from Borexino, KamLAND, and Super-Kamiokande. This provides a
complementary neutrino probe of the asteroid-mass PBH dark-matter window. Actually our constraints
will go slightly beyond the asteroid-mass window.

The rest of the paper is organized as follows: In \cref{sec:Source_Physics}, we discuss the physics at the 
source. We briefly review the superradiance mechanism and the production of a steady neutrino flux from the 
surrounding boson cloud. We then identify the relevant parameter space for asteroid-mass PBHs by imposing 
stability and lifetime conditions for both scalar and vector clouds, and compute the corresponding neutrino
flux at the source. In the following \cref{sec:Propagation}, we study the propagation of the neutrino signal 
from the source to the detector, including both Galactic and extragalactic PBH populations, for the allowed
scalar and vector parameter spaces. In \cref{sec:Constraints}, we describe our constraint procedure and
compare the predicted flux with existing low-energy neutrino bounds. Finally, in \cref{sec:Summary}, we 
summarize our results and conclude.

\section{Physics at the source}
\label{sec:Source_Physics}

As discussed above, we study the neutrino emission from PBHs surrounded by superradiantly produced
clouds of light bosons \cite{1971JETPL..14..180Z, 1971NPhS..229..177P, 1972JETP...35.1085Z}. In this
section, we focus on the source-level physics relevant for this signal. We first identify the parameter
space in which such clouds can form and remain stable over the age of the Universe, and then summarize
how their coupling to neutrinos leads to a steady neutrino flux. For the asteroid-mass PBHs considered
here, the resulting neutrinos can have energies of order tens of MeV or below. We derive the neutrino
energy and flux from a single PBH for both scalar and vector clouds, while the propagation to Earth
and the resulting constraints are discussed in the following sections.

Black-hole superradiance is a well-studied mechanism through which a rotating black hole can transfer
part of its angular momentum to a light bosonic field; see, e.g.,
Refs.~\cite{sr00Brito:2015oca, Zouros:1979iw, sr1Brito:2014wla} for a review. When the boson
Compton wavelength  ($\lambda_C= 1/m_\phi$)\footnote{
We use natural units here with $\hbar = c = 1$.
} is comparable to the black-hole horizon size ($r_g \equiv M_\text{BH}$), the field can form a gravitationally bound cloud around the black hole and grow through superradiant amplification. The efficiency of the superradiance process is commonly characterized by the gravitational fine-structure constant $\alpha_g$, defined as \cite{Arvanitaki:2016qwi}
\begin{align}
        \label{eq:SR_Condition}
       \ag \equiv G_N \, \mphi \, \mpbh \sim 0.75 \left(\frac{\mphi}{10^{-10} \text{ eV}}\right)\left(\frac{\mpbh}{M_{\odot}}\right)\lesssim 1 \;,
\end{align}
where $G_N$ is Newton's constant and $\mpbh$ is the PBH mass.

For a given value of $\alpha_g$, PBHs in the asteroid-mass range
$10^{17}$--$10^{23} \text{ g}$ can probe light bosons over a wide range of
masses. Using $M_\odot = 1.99\times 10^{33} \text{ g}$ for reference, the
boson mass can be written as
\begin{align}
    \label{eq:mphi_scaling}
    \mphi \simeq 2.7 \times 10^{2} \left( \frac{\alpha_g}{0.1} \right) \left( \frac{10^{20} \text{ g}}{\mpbh} \right) \text{ eV}.
\end{align}
Thus, for $\alpha_g=\mathcal{O}(10^{-3}-10^{-1})$, asteroid-mass PBHs are
sensitive to boson masses spanning roughly
$\mphi \sim 10^{-3} - 10^{5}$~eV.
Actually using these ranges for $\ag$
and $\mphi$
we will go somewhat beyond the asteroid mass range as will become
apparent later. In our numerical analysis, we scan over
these ranges of $\alpha_g$ and $\mphi$, which slightly extends the corresponding
PBH mass range beyond the nominal asteroid-mass window, as we will discuss
below.

Now, if the bosonic field around the PBH couples to fermions, in our case:
neutrinos, this will turn the spinning PBH into a fermion factory, as it was discussed
in great detail in \cite{Chen:2023vkq}. 
Following them, the average emitted neutrino energy from a scalar cloud is approximately
\begin{align}
    \label{eq:Neutrino_Energy_Scalar}
    \bar{E}_\nu^{s} \approx 2.7 \times 10^{10} \left(\frac{\gvp}{10^{-4}}\right) \left(\frac{\Psi_0}{10^{12} \text{ GeV}}\right) \text{ MeV}
    \;,
\end{align}
while for  a cloud composed  of vector bosons
\begin{align}
    \label{eq:Neutrino_Energy_Vector}
    \bar{E}_\nu^{v} \approx 3.5 \times 10^{14} \left(\frac{\gvV}{10^{-4}}\right) \left(\frac{\Psi_0}{10^{16} \text{ GeV}}\right) \text{ MeV}\;.
\end{align}
Here $\gvp$ is the Yukawa coupling of the scalar $\phi$ to the neutrino, and $\gvV$ is
the gauge coupling between the vector $V$ and the neutrino.
For simplicity, we assume that both couplings are diagonal and universal in the neutrino mass basis, so that all three neutrino mass eigenstates couple with the same strength. To distinguish the scalar and vector cases, we denote the vector-boson mass by $\mV$, while reserving $\mphi$ for the scalar-boson mass.

The average flux per neutrino flavor coming from a single spinning 
PBH, a scalar cloud evaluated at the average neutrino energy,
$\bar{E}_\nu^s$,
is given by \cite{Chen:2023vkq}\footnote{ While we 
follow \cite{Chen:2023vkq} as main reference,  we will adapt
the notation into our conventions. We also rewrite the expressions
into flux per neutrino flavor for both cases, vector and scalar.
So for a coupling to all three active neutrino generations the fluxes
must be multiplied with a factor of three.
}
\begin{align}
    \label{eq:Neutrino_Flux_From_One_Source_Scalar}
    \frac{\diff \Phi_\nu^{s}}{\diff E_\nu} (\bar{E}_\nu^{s}) &\approx
        \frac{1.1 \times 10^{-20}}{\text{ cm}^2 \text{ s} \text{ eV} } 
        \left( \frac{\Psi_0}{4.8 \times 10^7 \text{ GeV}} \right)^{1/2} 
        \nonumber\\
        &\phantom{\approx} \times
        \left( \frac{1 \text{ eV}}{m_\phi} \right)^{1/2}
        \left( \frac{0.1}{\alpha_g} \right)^3
        \left( \frac{\gvp}{10^{-4}}\right)^{1/2}
        \left( \frac{5 \text{ kpc}}{d}\right)^{2} 
        \;,
\end{align}
while for the vector case
\begin{align}
    \label{eq:Neutrino_Flux_From_One_Source_Vector}
    \frac{\diff \Phi_\nu^{v}}{\diff E_\nu} (\bar{E}_\nu^{v}) &\approx
        \frac{3.5 \times 10^{-9}}{\text{ cm}^2 \text{ s} \text{ eV} } 
        \left( \frac{\Psi_0}{5.7 \times 10^{14} \text{ GeV}} \right)
        \nonumber\\
        &\phantom{\approx} \times
        \left( \frac{1 \text{ eV}}{m_V} \right)
        \left( \frac{0.1}{\alpha_g} \right)^3
        \left( \frac{\gvV}{10^{-4}}\right)
        \left( \frac{5 \text{ kpc}}{d}\right)^{2} \;,
\end{align}
where $d$ is the distance between the observer and the PBH,
and $\Psi_0$ is the value of the bosonic field.

For the emitted neutrino energies and fluxes we will only take into account the average
values here. In reality, this would depend, for instance, on the emission direction. However,
since we will later integrate over a large population of PBHs, assuming a random distribution of
orientations, these angular variations are expected to average out. Moreover, the spread
in emitted neutrino energy is not very large, which allows us to treat the fluxes as effectively
monochromatic,
\begin{equation}
    \label{eq:Phi_0}
     \frac{\diff \Phi_\nu}{\diff E_\nu} (\bar{E}_\nu) \equiv \Phi_0(\bar{E}_\nu) \, \delta(\bar{E}_\nu-E_\nu) \;,
\end{equation}
such that when we integrate the right-hand side over the neutrino energy we get back the
$\diff \Phi_\nu/\diff E_\nu$ from above.

Here we consider a scenario in which the bosonic cloud grows until it reaches an equilibrium.
At this stage, the  energy  carried away by neutrino emission  balances the energy extracted
from the PBH through superradiance, leading to a  steady and continuous neutrino flux from the
cloud. We neglect other possible sources of energy gains and loss (such as emission into other
particle species, accretion of ordinary matter,
merging of PBHs, etc.). For such a saturated configuration to be realized, the parameter space
of the scalar and the PBH must satisfy the two conditions 
\begin{align}
    \label{eq:Stability_Conditions_Scalar}
    \Psi_0^{c,s} < \Psi_0^{\text{max},s} \quad \text{ and } \quad 
    \gvp  \, \Psi_0^{c,s} > m_\nu \;,
\end{align}
where we take the neutrino mass $m_\nu = 0.1$~eV and the maximum allowed  field amplitude is 
\begin{equation}
    \Psi_0^{\text{max},s} = 2.8 \times 10^{15} \left(\frac{\alpha_g}{0.1}\right)^2 \sqrt{\frac{M_B/\mpbh}{0.1}} \text{ GeV,}
\end{equation}
where $M_B$ denotes the total mass of the bosonic cloud and  the ratio $M_B/\mpbh$ specifies
the fraction of the PBH mass transferred to the cloud during the superradiant growth.

For a vector case, the first condition remains unchanged, but the kinematic condition for
neutrino production  has to be modified and we have the following  conditions
\begin{equation}
    \label{eq:Stability_Conditions_Vector}
    \Psi_0^{c,v} < \Psi_0^{\text{max},v} \quad \text{ and } \quad 
    m_V\,\gvV \, \Psi_0^{c,v} > m_{\nu}^2 \text{,}
\end{equation}
and
\begin{equation}
    \Psi_0^{\text{max},v} = 2.2 \times 10^{16} \left(\frac{\alpha_g}{0.1}\right)^2 \sqrt{\frac{M_B/\mpbh}{0.1}} \text{ GeV.}
\end{equation}
Here $\Psi_0^c$ denotes the critical field amplitude at which the energy
loss into neutrinos balances the superradiant energy gain in terms of the cloud growth. For a scalar
cloud, this critical amplitude is given by
\begin{equation}
    \label{eq:Psi0c_Scalar}
    \Psi_0^{c,s} = 1.2 \times 10^{-4} N_\nu^2
    \left(\frac{m_\phi}{1 \text{ eV}}\right) \left(\frac{\alpha_g}{0.1}\right)^{16} \left(\frac{\tilde{a}}{0.9}\right)^2 \left(\frac{10^{-4}}{\gvp}\right)^5 \text{ GeV,}
\end{equation}
and for a vector cloud
\begin{equation}
    \label{eq:Psi0c_Vector}
    \Psi_0^{c,v} = 7.8 \times 10^{-1} N_\nu
    \left(\frac{m_V}{1 \text{ eV}}\right) \left(\frac{\alpha_g}{0.1}\right)^{6} \left(\frac{\tilde{a}}{0.9}\right) \left(\frac{10^{-4}}{\gvV}\right)^3 \text{ GeV,}
\end{equation}
at which the cloud is stable, see \cite{Chen:2023vkq}. 
Here $\tilde{a}$  is the initial value of the dimensionless spin parameter of the PBH.
The angular momentum of the PBH changes over time as rotational energy is extracted but the
mass of the bosonic cloud and the critical field value and related observables are to a
good approximation constant and only depend on an initial condition which we treat as a
free parameter.
We also note that, in particular, for the scalar
case the prefactor is different than in \cite{Chen:2023vkq} since
we present our results per neutrino flavor. 
For clarity, we label the number of flavors here with $N_\nu$
in the following.

Let us have a closer look at these two constraints and we begin with the scalar case.
The first requirement,
\begin{equation}
    \label{eq:Constraint_Max_Field_Value}
    \Psi_{0}^{c,s} < \Psi_{0}^{\text{max},s} \;,
\end{equation}
enforces that the critical scalar field amplitude necessary for efficient 
neutrino production does not exceed the maximum field amplitude attainable 
via PBH superradiance. 
This condition imposes a lower bound on the coupling $\gvp$, since for sufficiently 
small $\gvp$ the critical field amplitude becomes too large to be attained by the cloud.
This lower bound can be made explicit by substituting the expressions for $\Psi_{0}^{c,s}$
and $\Psi_{0}^{\text{max},s}$ into \cref{eq:Constraint_Max_Field_Value}
\begin{align}
    \label{eq:Stability_Condition1_Scalar_Explicit}
    4.4 \times 10^{-20} N_\nu^2 \left(\frac{m_\phi}{1 \text{ eV}} \right) \left(\frac{\alpha_g}{0.1}\right)^{14} \left(\frac{\tilde{a}}{0.9}\right)^2 \sqrt{\frac{0.1}{M_B/\mpbh}} 
    \lesssim
    \left(\frac{\gvp}{10^{-4}}\right)^{5}  \;.
\end{align}
The right-hand side goes with the fifth power of $\gvp$ so to lower
the bound on it the left-hand side has to change drastically as well. But
as we can see the dependence on $m_\phi$ and $\tilde{a}$ are rather weak.
We can expect the largest  from $\ag$ but as we will discuss soon there is
another constraint taken into account which is usually stronger than this
constraint. The bound depends on $\gvp$ through the fifth power, so lowering the
required value of $\gvp$ would require a substantial change in the remaining parameters.
The dependence on $m_\phi$ and $\tilde{a}$ is rather weak comparatively. The strongest
parametric effect can therefore come from $\alpha_g$. As we will discuss below, however,
the second condition typically provides the stronger constraint on the allowed parameter space.

For the vector case we can do the same exercise and rewrite 
the first condition of \cref{eq:Stability_Conditions_Vector} 
in the same way
\begin{align}
    \label{eq:Stability_Condition1_Vector_Explicit}
    3.6 \times 10^{-17} N_\nu \left(\frac{\mV}{1 \text{ eV}} \right) \left(\frac{\alpha_g}{0.1}\right)^{4} \left(\frac{\tilde{a}}{0.9}\right) \sqrt{\frac{0.1}{M_B/\mpbh}} 
    \lesssim
    \left(\frac{\gvV}{10^{-4}}\right)^{3} \;.
\end{align}

The second consistency condition for the scalar cloud,
\begin{equation}
	\gvp \, \Psi_{0}^{c,s} > m_\nu \;,
\end{equation}
translates more explicitly into
\begin{align}
    \label{eq:Upper_Bound_gvp_Scalar}
     1.2 \times 10^{2} N_\nu^2  \left(\frac{m_\phi}{1 \text{ eV}}\right) \left( \frac{0.1 \text{ eV}}{m_\nu} \right) \left(\frac{\alpha_g}{0.1}\right)^{16} \left(\frac{\tilde{a}}{0.9}\right)^2 
     \gtrsim  \left(\frac{\gvp}{10^{-4}}\right)^4 \;.
     \;
\end{align}
This condition provides an upper bound on $\gvp$, although in practice this bound is
quite large and often lies above the perturbative regime.
In our numerical analysis, we nevertheless impose this condition, while also restricting
the scan ranges to $\gvp \leq 1$ and $\gvV \leq 1$.

For the vector case, the second consistency is a bit different
from the scalar one,
\begin{equation}
    \mV \, \gvV \, \Psi_0^{c,v} > m_{\nu}^2 \; ,
\end{equation}
and the explicit form is
\begin{align}
    \label{eq:Upper_Bound_gvp_Vector}
    7.8 \times 10^6 \,  N_\nu \left(\frac{\mV}{1 \text{ eV}}\right)^{2} \left( \frac{0.1 \text{ eV}}{m_\nu} \right)^{2} \left(\frac{\alpha_g}{0.1}\right)^{6} \left(\frac{\tilde{a}}{0.9}\right) 
    \gtrsim  \left(\frac{\gvV}{10^{-4}}\right)^2
     \;.
\end{align}

Let us now discuss the lifetime constraints. As discussed above, we work in the so-called saturation 
or balanced phase where the bosonic cloud is approximately stable. The energy carried away by neutrino
emission is compensated by the rotational energy of
the PBH, which is finite. This therefore introduces an additional timescale that must be taken into account.

The duration of the saturation (balanced) phase can be estimated as (see, for instance, \cite{Chen:2023vkq})
\begin{equation} 
    \tau_{\text{sat}} \sim \frac{\mpbh}{\Gamma_{\text{SR}} \, M_B} \; .
    \label{eq:tau_sat}
\end{equation}
It is the time-scale within which the spin of the black hole reaches the critical
value $\tilde{a}_{\text{crit}}$, exhausting the superradiance condition.
For a scalar cloud, the superradiant growth rate $\Gamma_{\text{SR}}$ is
\begin{align}
    \Gamma_{\text{SR}}^s \sim  \frac{\ag^8 \, \atilde \,  \mphi}{24} \;,
\end{align}
while for a vector boson cloud
\begin{align}
    \Gamma_{\text{SR}}^v \sim 4 \, \ag^6 \, \atilde \,m_V\;,
\end{align}
with  $m_V$ denoting the vector-boson mass.
The product $\Gamma_{\text{SR}} \, M_B$ thus represents the linear rate at which energy is 
extracted from the black hole via superradiance due to the bosonic particle. 
Here $\tilde{a}$ denotes the initial spin parameter of the PBH, which acts as a constant when evaluating the 
saturated configuration. It does not denote the time-dependent spin during the spin-down process.

During the saturation phase, the bosonic cloud has the critical amplitude $\Psi_0^{c}$. At this point,
the mass stored in the scalar cloud is given by \cite{Chen:2023vkq}
\begin{align}
    M_B^s = \frac{186 \, (\Psi_0^{c,s})^2}{\ag^3 \, \mphi} \;,
\end{align}
while for  the vector cloud, one has 
\begin{align}
    M_B^v = \frac{\pi \, (\Psi_0^{c,v})^2}{\ag^3 \, m_V } \;.
\end{align}

For the neutrino signal considered here to be realized, the bosonic cloud must
be in
the saturated, or balanced, phase for the age of the Universe. We therefore impose
\begin{align}
    \label{eq:Age_Constraint}
    \tau_{\text{sat}} \gtrsim T_{\text{univ}} \;,
\end{align}
where $T_{\text{univ}} = 1.4\times 10^{10}$~years
is the age of the universe. 
This condition further ensures that  the PBHs considered here can emit a steady and continuous flux
of neutrinos from the very early universe until today. We now write this condition more explicitly,
starting with the scalar case, 
\begin{align}
    \label{eq:Lifetime_Constraint_Scalar}
    T_{\text{univ}} &\lesssim \frac{\mpbh}{\Gamma_{\text{SR}}^s \, M_B^s} \nonumber\\
    &\Rightarrow 4.7 \times 10^{-17} N_\nu^4 
    \left(\frac{m_\phi}{1 \text{ eV}}\right)^3
    \left(\frac{\alpha_g}{0.1}\right)^{36} 
    \left(\frac{\tilde{a}}{0.9}\right)^5  \lesssim \left(\frac{\gvp}{10^{-4}}\right)^{10} \;,
\end{align}
and for the vector case
\begin{align}
    \label{eq:Lifetime_Constraint_Vector}
    T_{\text{univ}} &\lesssim \frac{\mpbh}{\Gamma_{\text{SR}}^v \, M_B^v} \nonumber\\
    &\Rightarrow 3.1 \times 10^{-7} N_\nu^2 
    \left(\frac{m_V}{1 \text{ eV}}\right)^3 
    \left(\frac{\alpha_g}{0.1}\right)^{14} 
    \left(\frac{\tilde{a}}{0.9}\right)^3  \lesssim \left(\frac{\gvV}{10^{-4}}\right)^{6} \;.
\end{align}

\begin{figure}
    \centering
    \includegraphics[width=\textwidth]{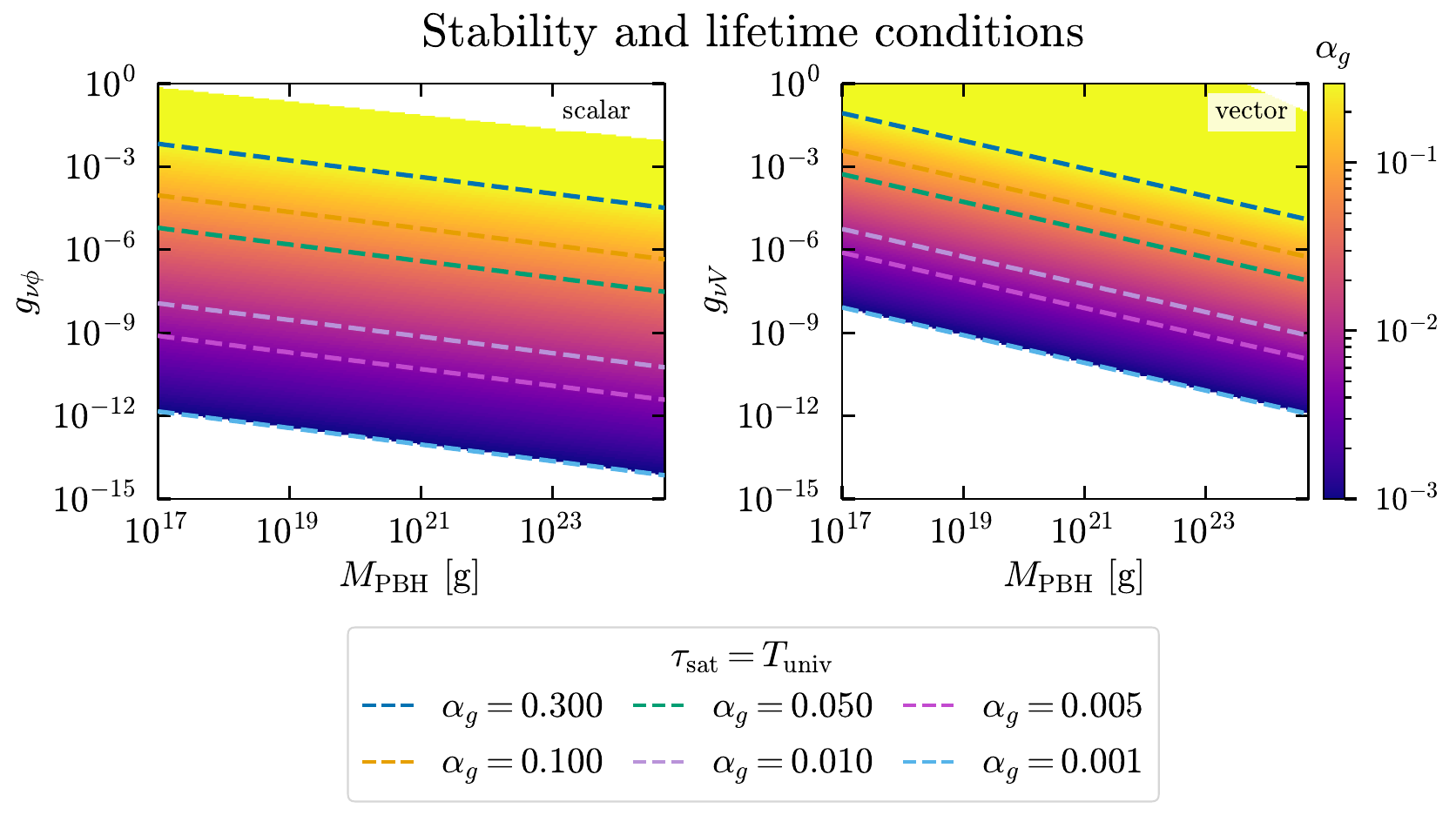}
    \caption{Here we sketch the allowed parameter space for the PBH mass, $\mpbh$, gravitational coupling constant $\ag$ and the scalar-neutrino coupling, $\gvp$, and the
    vector-neutrino coupling, $\gvV$, respectively.
    We impose the stability constraints \cref{eq:Stability_Conditions_Scalar},
    \cref{eq:Stability_Conditions_Vector} and the age constraint
    \cref{eq:Age_Constraint}. These constraints make sure that the PBHs are to a good
    approximation emitting a steady flux of neutrinos throughout the age of the universe.
    The forbidden regions are white, while the colored regions show the maximum 
    allowed gravitational coupling constant $\ag$. The maximal allowed value overall in our
    scans was about 0.4. 
    Dashed lines indicate the combinations of $(\mpbh, \gvp)$ for which
    the saturation timescale $\tau_{\text{sat}}$ equals the age of the 
    universe ($T_{\text{univ}}$) at fixed values of $\ag$, such that the region
    below each contour corresponds to $\tau_{\text{sat}} < T_{\text{univ}}$.
    In both plots we set $\atilde = 0.9$ and
    $N_\nu = 1$. For more details, see main text.
    \label{fig:Parameters_Continuous_Flow}
    }
\end{figure}

In \cref{fig:Parameters_Continuous_Flow} we show the parameter space allowed by the constraints from
\cref{eq:Stability_Conditions_Scalar}, \cref{eq:Stability_Conditions_Vector} and \cref{eq:Age_Constraint}.
The white regions are excluded, while the colored regions correspond to the allowed parameter space.
The color bar indicates the value of the gravitational fine-structure constant $\alpha_g$. We find that viable
parameter space exists  across the full asteroid-mass range of PBHs. The couplings $\gvp$ and $\gvV$ can also vary
over a wide range, although the lower bound on $\gvV$ is noticeably stronger than in the scalar case.
Similarly, $\alpha_g$ can span a broad range, from below $10^{-3}$ up to values of order unity. The dashed
contours indicate the combinations of $(\mpbh,\gvp)$ and $(\mpbh,\gvV)$ for which the saturation timescale
$\tau_{\text{sat}}$ equals the age of the Universe, $T_{\text{univ}}$, for fixed values of $\alpha_g$.
Regions below each contour correspond to $\tau_{\text{sat}}<T_{\text{univ}}$, implying that the cloud
can be in the saturated phase within cosmic time. The careful reader will
notice that in \cref{fig:Parameters_Continuous_Flow} we plotted  $\mpbh$ beyond the asteroid-mass
range. We do this to show here the full region for which we can derive
a constraint.

We now rewrite the expressions for the average neutrino energy and the source flux in order to
derive the corresponding bounds on these quantities. These bounds follow from the stability and
lifetime conditions discussed above.

We first consider the scalar case. Substituting the $\Psi_0$ with  critical field amplitude
$\Psi_0^{c,s}$ from \cref{eq:Psi0c_Scalar} we find
\begin{align}
    \label{eq:Neutrino_Energy_Scalar_no_PSi0}
    \bar{E}_\nu^s \approx 3.3  \times 10^{-6} N_\nu^2 \left(\frac{m_\phi}{1 \text{ eV}}\right) \left(\frac{\alpha_g}{0.1}\right)^{16} \left(\frac{\tilde{a}}{0.9}\right)^2  \left(\frac{10^{-4}}{\gvp}\right)^4  \text{ MeV,}
\end{align}
and for the vector one, we replace $\Psi_0$ with the critical field density $\Psi_0^{c,v}$ from \cref{eq:Psi0c_Vector} of the vector case
\begin{align}
    \label{eq:Neutrino_Energy_Vector_no_PSi0}
    \bar{E}_\nu^v \approx 2.7  \times 10^{-2} N_{\nu}\left(\frac{m_V}{1 \text{ eV}}\right) \left(\frac{\alpha_g}{0.1}\right)^{6} \left(\frac{\tilde{a}}{0.9}\right)  \left(\frac{10^{-4}}{\gvV}\right)^2  \text{ MeV.}
\end{align}

Now we use the stability and lifetime constraints to get upper
and lower bounds for the neutrino energies in both cases.
For the scalar case, using \cref{eq:Upper_Bound_gvp_Scalar}
and \cref{eq:Lifetime_Constraint_Scalar}, we obtain
\begin{align}
    \label{eq:Enus_Bounds}
    \bar{E}_\nu^s \begin{cases}
        \lesssim  11 \,  N_\nu^{2/5} 
        \left(\frac{1 \text{ eV}}{m_\phi} \right)^{1/5} \left(\frac{\alpha_g}{0.1}\right)^{8/5} 
        \text{ MeV,}  \\
        \gtrsim 0.27 \, m_\nu 
        = 2.7 \times 10^{-8} \left(\frac{m_{\nu}}{0.1\text{ eV}}\right) \text{ MeV,}
    \end{cases}
\end{align}
and for the vector case using \cref{eq:Upper_Bound_gvp_Vector} and \cref{eq:Lifetime_Constraint_Vector}, we find
\begin{align}
    \label{eq:Enuv_Bounds}
    \bar{E}_\nu^v \begin{cases}
        \lesssim 3.9 \, N_\nu^{1/3} 
        \left(\frac{\alpha_g}{0.1}\right)^{4/3} 
        \text{ MeV,}  \\
        \gtrsim  3.5  \times 10^{-9} 
        \left(\frac{1 \text{ eV}}{m_V}\right) 
        \left(\frac{m_\nu}{0.1 \text{ eV}}\right)^{2} 
        \text{ MeV.}  
    \end{cases}
\end{align}

We now turn to the source flux. For the scalar case of the neutrino flux per neutrino flavor
we find after 
replacing $\Psi_0$ in \cref{eq:Neutrino_Flux_From_One_Source_Scalar} 
with $\Psi_0^{c,s}$ for $\Phi_0^s$
\begin{align}
    \label{eq:Neutrino_Flux_From_One_Source_Scalar_no_Psi0}
    \Phi_0^s
    &\approx
        \frac{1.7 \times 10^{-20}}{\text{ cm}^2 \text{ s} }
        N_\nu
        \left( \frac{\alpha_g}{0.1} \right)^5
        \left(\frac{\tilde{a}}{0.9}\right)
        \left(\frac{10^{-4}}{\gvp}\right)^{2} 
        \left( \frac{5 \text{ kpc}}{d}\right)^{2} 
        \;.
\end{align}
This also has an upper and lower bound
\begin{align}
    \label{eq:Neutrino_Flux_From_One_Source_Scalar_Bounds}
    \Phi_0^s
    &
    \begin{cases} 
        \lesssim
        \frac{3.2 \times 10^{-17}}{\text{ cm}^2 \text{ s} }
        N_\nu^{1/5}
        \left( \frac{1 \text{ eV}}{m_\phi} \right)^{3/5}
        \left( \frac{0.1}{\alpha_g} \right)^{11/5}
        \left( \frac{5 \text{ kpc}}{d}\right)^{2}
        \;,\\
        \gtrsim
        \frac{1.6 \times 10^{-21}}{\text{ cm}^2 \text{ s} }
        \left( \frac{1 \text{ eV}}{m_\phi} \right)^{1/2}
        \left( \frac{m_\nu}{0.1 \text{ eV}} \right)^{1/2}
        \left( \frac{0.1}{\alpha_g} \right)^{3}
        \left( \frac{5 \text{ kpc}}{d}\right)^{2} 
        \;.\\
    \end{cases} 
\end{align}

And in the same way we find for the vector case after replacing
$\Psi_0$ in in \cref{eq:Neutrino_Flux_From_One_Source_Vector} 
with $\Psi_0^{c,v}$ for $\Phi_0^v$
\begin{align}
    \label{eq:Neutrino_Flux_From_One_Source_Vector_no_Psi0}
    \Phi_0^v    
    &\approx
        \frac{4.8 \times 10^{-18}}{\text{ cm}^2 \text{ s} }
        N_\nu
        \left( \frac{\alpha_g}{0.1} \right)^{3}
        \left(\frac{\tilde{a}}{0.9}\right)
        \left(\frac{10^{-4}}{\gvV}\right)^2 
        \left( \frac{5 \text{ kpc}}{d}\right)^{2} \;.
\end{align}
The vector flux is also bounded from above and below. Using the corresponding stability
and lifetime constraints, we find
\begin{align}
    \label{eq:Neutrino_Flux_From_One_Source_Vector_Bounds}
    \Phi_0^v  
    &
    \begin{cases} 
        \lesssim
        \frac{6.9 \times 10^{-17}}{\text{ cm}^2 \text{ s} }
        N_\nu^{1/3}
        \left( \frac{1\text{ eV}}{m_V}  \right)
        \left( \frac{0.1}{\alpha_g} \right)^{5/3}
        \left( \frac{5 \text{ kpc}}{d}\right)^{2} 
        \;,\\
        \gtrsim
        \frac{6.2 \times 10^{-25}}{\text{ cm}^2 \text{ s} }
        \left( \frac{m_\nu}{0.1 \text{ eV}} \right)^{2}
        \left( \frac{1 \text{ eV}}{m_V} \right)^{2}
        \left( \frac{0.1}{\alpha_g} \right)^{3}
        \left( \frac{5 \text{ kpc}}{d}\right)^{2} 
         \;.
    \end{cases} 
\end{align}
In both the scalar and vector cases, the upper bound on the flux is correlated with the 
upper bound on the emitted neutrino energy, while the lower bound on the flux is correlated
with the corresponding lower bound on the emitted energy.

\begin{figure}
    \centering
    \includegraphics[width=\linewidth]{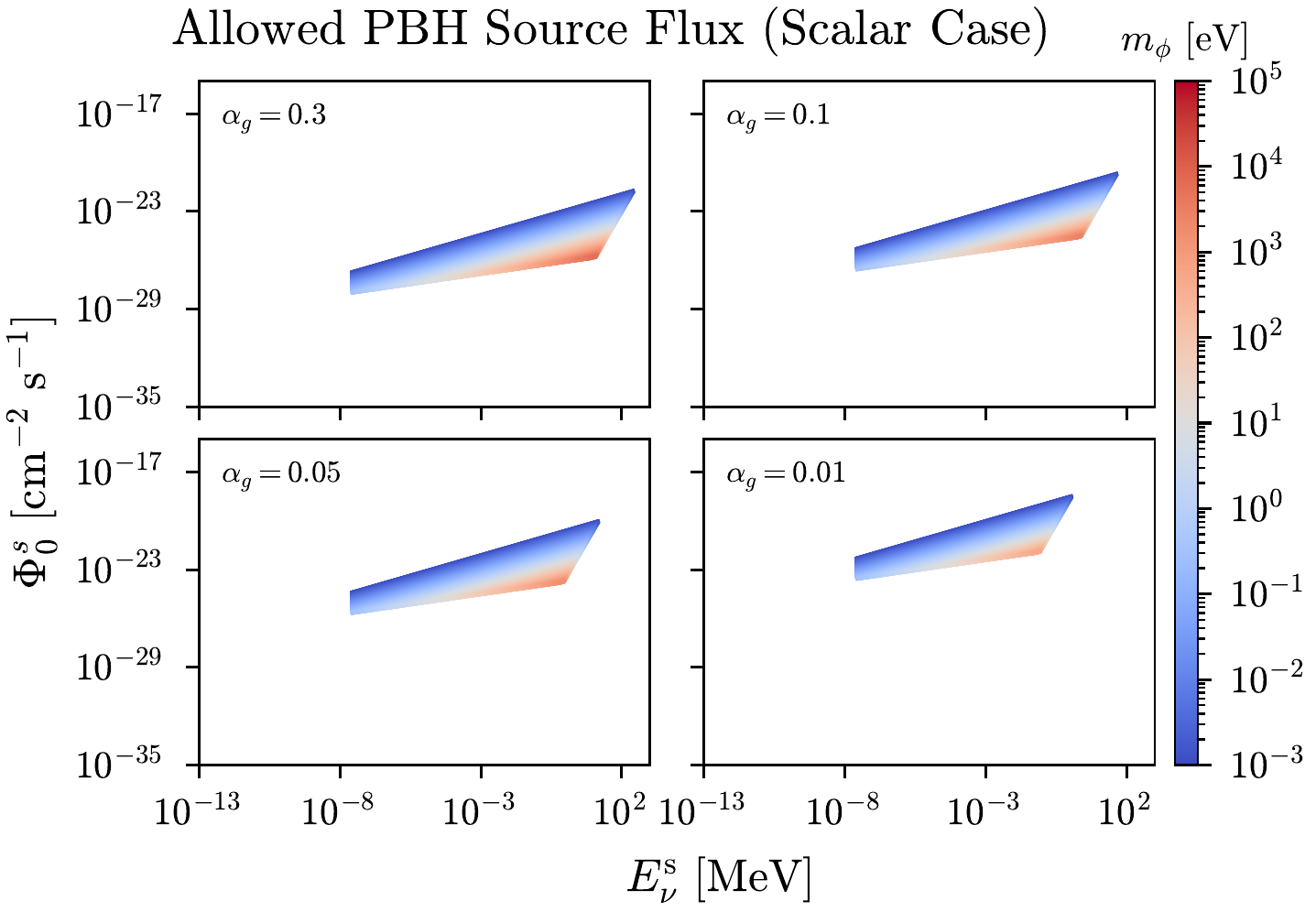}
    \caption{Allowed neutrino flux $\Phi_0^s$  as a function of neutrino energy $E_\nu^s$ for the scalar case. 
    Each panel shows the allowed flux-energy region for different $\ag$ values with the color scale
    representing the scalar mass $\mphi$.
    Here we have always set $d = 5$~kpc, $\tilde{a} = 0.9$ and $N_\nu = 1$.
    }
    \label{fig:FluxEnergy_scalar_mphi_lines}
\end{figure}

\begin{figure}
     \centering
     \includegraphics[width=\linewidth]{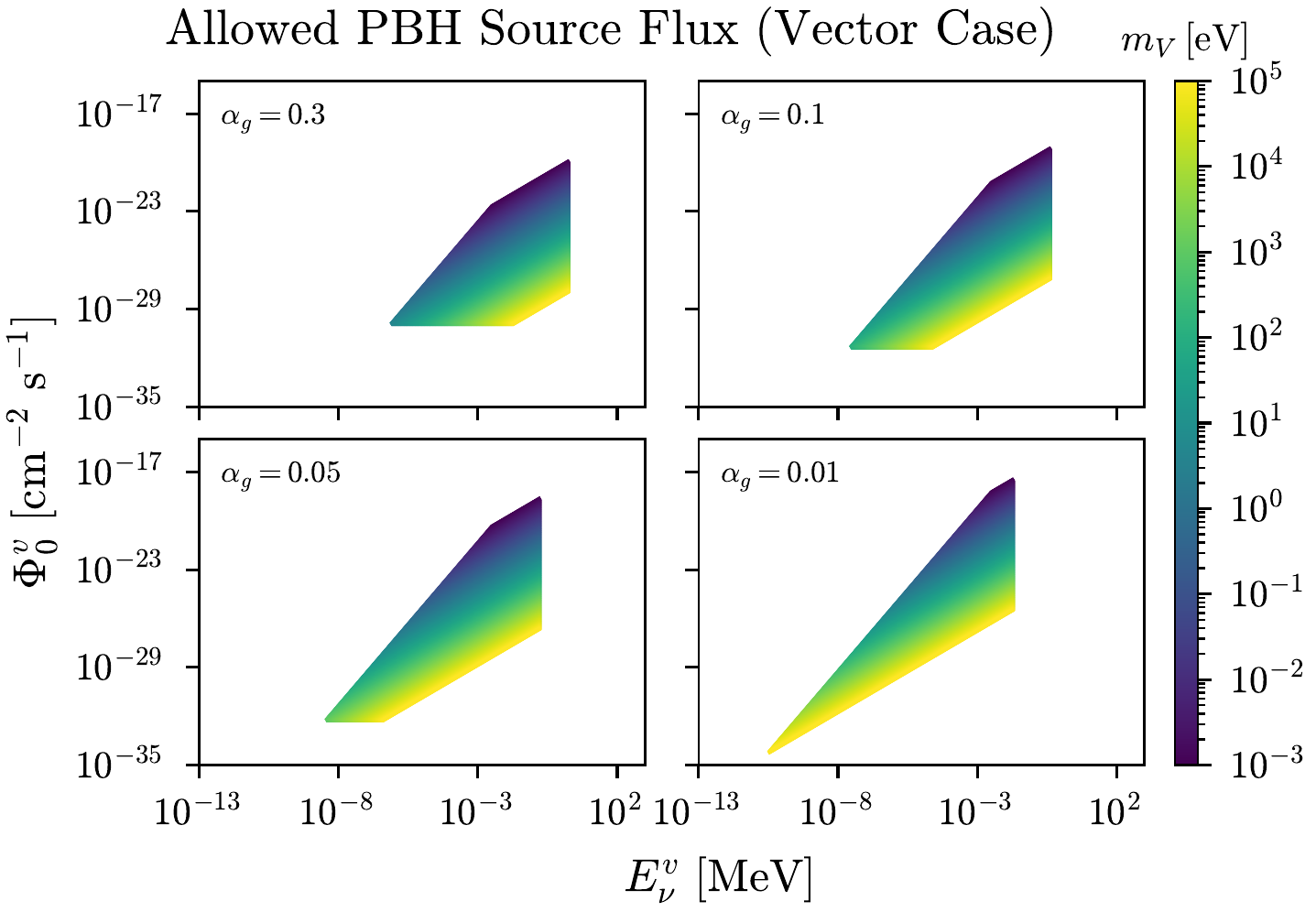}
     \caption{Allowed neutrino flux $\Phi_0^v$ as a function of neutrino energy $E_\nu^v$ for the vector case.
     Each panel shows the allowed flux-energy region for different $\ag$ values with the color scale
     representing the vector mass $\mV$.
     We set $d = 5$~kpc, $\tilde{a} = 0.9$ and $N_\nu = 1$ throughout.
     Compared to the scalar case in \cref{fig:FluxEnergy_scalar_mphi_lines} the upper bound on the
     neutrino energy is a bit lower but the flux itself can be a bit higher.
     }
    \label{fig:FluxEnergy_vector_mV_lines}
\end{figure}

We show examples of the allowed regions in the flux--energy plane in
\cref{fig:FluxEnergy_scalar_mphi_lines,fig:FluxEnergy_vector_mV_lines}
for the scalar and vector cases, respectively. In both figures we have set $d = 5$~kpc,
$\tilde{a} = 0.9$ and $N_\nu = 1$. The scalar case is more
restrictive in the sense that the allowed regions are generally smaller,
with a significantly narrower range of allowed fluxes. However, the maximum
allowed neutrino energy is higher in the scalar case, reaching up to tens of
MeV, whereas in the vector case it remains at the level of a few MeV. This
behavior is consistent with the analytic bounds in
\cref{eq:Enus_Bounds,eq:Enuv_Bounds}.

Having established the conditions under which a single asteroid-mass PBH can
produce a steady neutrino signal, together with the corresponding flux and
energy ranges, we now turn to the propagation of this signal and its
contribution to the neutrino flux observed at Earth.

\section{Propagation from Source to Detector}
\label{sec:Propagation}

As discussed in the previous section, PBHs surrounded by superradiant boson clouds can enter
a quasi-equilibrium state in which the cloud emits a neutrino flux with approximately constant
properties. If such PBHs are primordial and constitute a fraction of the dark matter, the observable
neutrino flux at Earth receives contributions from both the Galactic PBH population and the
extragalactic PBH population.

In this work, we focus on asteroid-mass PBHs in the range
$10^{17}\text{ g} < \mpbh < 10^{23} \text{ g}$. This mass window remains one of the least constrained
regions of PBH parameter space, where PBHs may still constitute all of the dark matter, corresponding
to $\fpbh=1$. In the previous section, we identified the conditions under which a single PBH can emit
a steady neutrino signal and derived the corresponding source-level flux and energy. We now use these
results to compute the neutrino flux reaching Earth from the PBH dark-matter population. This
calculation can be separated into two components: the extragalactic contribution and the Galactic
contribution.

\subsection{Extragalactic Flux}
\label{subsec:egflux}

Let $Q_\nu(\bar{E}_\nu)$ denote the differential neutrino emission rate of a single source, 
per unit energy and time, at redshift $z$.  The diffuse extragalactic neutrino flux, including
the contribution from all PBHs along the line of sight, can then be written as \cite{Ahlers:2018fkn}
\begin{equation}
    \frac{\diff \Phi_{\text{EG}}}{\diff E_\nu \diff \Omega} = \frac{c}{4\pi} \int_0^{z_{\max}} \diff z \; \frac{n_0}{H(z)}
Q_\nu \big((1+z)E_\nu\big) \;,
    \label{eq:Ahlers_general}
\end{equation}
where we introduce the Hubble parameter as a function of redshift
\begin{equation}
    H(z) = H_0 \sqrt{\Omega_m (1+z)^3 + \Omega_r (1+z)^4 + \Omega_\Lambda} \;,
\end{equation}
with $H_0 = 67.66$~km/(s Mpc), $\Omega_m = 0.31$, 
$\Omega_r h^2 = 2.47\times 10^{-5}$ and 
$\Omega_\Lambda = 0.69$~\cite{Planck:2018vyg}. Also we have $E_\nu$ as the neutrino energy observed on Earth,
while $\bar{E}_\nu=(1+z)E_\nu$ is the corresponding energy at the source.
The comoving number density of PBHs is given as
\begin{align}
    n_0 &= \frac{\rho_{\text{DM},0} \,  \fpbh} {\mpbh} = \frac{2.24\times 10^{-47}}{\text{cm}^3} \frac{\fpbh}{1.0}\frac{10^{17} \text g}{\mpbh}  \;,
\end{align}
where we used the following numerical values \cite{Planck:2018vyg}
\begin{equation}
\begin{split}
    \rho_{\text{DM},0} &= \Omega_{\text{DM}} \, \rho_{c,0} = 2.24\times 10^{-30}   \text{ g cm}^{-3} \;, \\
    \rho_{c,0} &= 1.054 \times 10^{-5} h^2  \frac{\text{GeV}}{\text{cm}^3}  \;, \\
    h &= 0.674 \,, \quad \Omega_{\text{DM}} = 0.265\;.
\end{split}    
\end{equation}

The boson-BH system considered in \cite{Chen:2023vkq,Banerjee:2024knt, Banerjee:2025ddr}
emits neutrinos which are rather sharply peaked at $\bar{E}_\nu$, 
given in \cref{eq:Neutrino_Energy_Scalar_no_PSi0} for the scalar case
and in \cref{eq:Neutrino_Energy_Vector_no_PSi0} for the vector case.
We can therefore model the neutrino emission rate to a good approximation
in terms of a delta function
\begin{equation}
    \label{eq:Q_nu}
    Q_\nu(\bar{E}_\nu) = Q_0(\bar{E}_\nu) \, \bar{E}_\nu \, \delta(\bar{E}_\nu-E_\nu) \;,
\end{equation}
where
\begin{equation}
    \label{eq:Q0}
    Q_0(\bar{E}_\nu) = 4 \, \pi \, d^2  \frac{\Phi_0}{\bar{E}_\nu} \;, 
\end{equation}
and the fluxes for the considered cases are given in
\cref{eq:Neutrino_Flux_From_One_Source_Scalar_no_Psi0} and
\cref{eq:Neutrino_Flux_From_One_Source_Vector_no_Psi0}. 
Note that there is an indirect $z$-dependence via the dependence
on $\bar{E}_\nu$.

\begin{figure}
    \centering
    \includegraphics[width=0.6\linewidth]{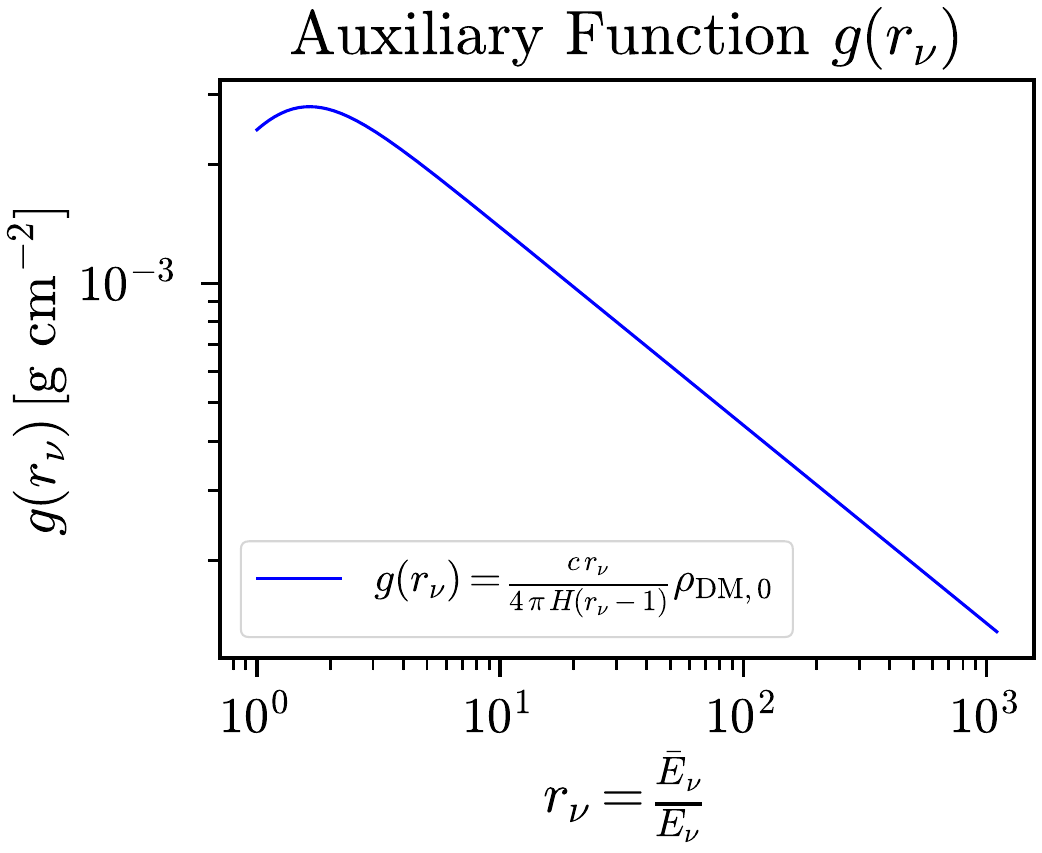}
    \caption{Universal auxiliary function which appears
    in our expression for the extragalactic flux in 
    \cref{eq:Extragalactic_flux_at_Observed_Energy}. This part
    does not depend on any PBH parameters and it just multiplies
    the function $Q_0$ which contains all the information on the
    sources.
    }
    \label{fig:EG_Universal_Prefactor}
\end{figure}

Inserting \cref{eq:Q_nu} into \cref{eq:Ahlers_general},
\begin{equation}
    \frac{\diff \Phi_{\text{EG}}}{\diff E_\nu \diff \Omega} = \frac{c}{4\pi} \int \diff z \, \frac{n_0 \, Q_0( (1+z) E_\nu) \, \bar{E}_\nu}{H(z)} \delta\left( (1+z) E_\nu - \bar{E}_\nu \right) \;, 
\end{equation}
and using the identities
\begin{equation}
    \label{eq:delta_identity}
    \delta \left( (1+z) E_\nu - \bar{E}_\nu \right) = \frac{1}{E_\nu} \delta\!\left(z - z_*\right) \;,
    \qquad
    z_* = \frac{\bar{E}_\nu}{E_\nu}-1 \;,
\end{equation}
it is easy to evaluate the integral and we find
for the  extragalactic flux at observed energy $E_\nu$
\begin{align}
    \label{eq:Extragalactic_flux_at_Observed_Energy}
    \frac{\diff \Phi_{\text{EG}}}{\diff E_\nu \diff \Omega}
    &=
    \frac{c \, \bar{E}_\nu}{4 \, \pi \, E_\nu} \frac{n_0 \, Q_0(
    \bar{E}_\nu)}{H(\bar{E}_\nu/E_\nu -1 )} 
    = \frac{\fpbh} {\mpbh} g(r_\nu) \, Q_0(\bar{E}_\nu)  \;, \qquad
    E_\nu \le \bar{E}_\nu  \;,
\end{align}
where we have defined
\begin{align}
    \label{eq:auxiliary_fun}
    g(r_\nu) \equiv \frac{c \, r_\nu}{4 \, \pi}\frac{\rho_{\text{DM},0}}{H(r_\nu - 1)}  \quad \text{and} \quad r_\nu = \bar{E}_\nu/E_\nu \;.
\end{align}

The function $g(r_\nu)$ is universal, but it does not follow a simple power-law scaling.
Once the model parameters are fixed, the extragalactic flux can therefore be obtained by
multiplying $(\fpbh/\mpbh) \, Q_0(\bar{E}_\nu)$ by the corresponding value of the universal
function $g(r_\nu)$ shown in \cref{fig:EG_Universal_Prefactor}. In our calculation, we
integrate over redshift from $z=0$ up to $z=1100$, corresponding approximately to the
CMB epoch. In principle, the upper limit could be extended to the PBH formation epoch or
to neutrino decoupling. However, the contribution from very large redshifts is strongly
suppressed, as shown in \cref{fig:EG_Universal_Prefactor}. Our results are therefore only
weakly sensitive to the precise choice of the upper integration limit.

\begin{figure}
    \centering
    \includegraphics[width=\linewidth]{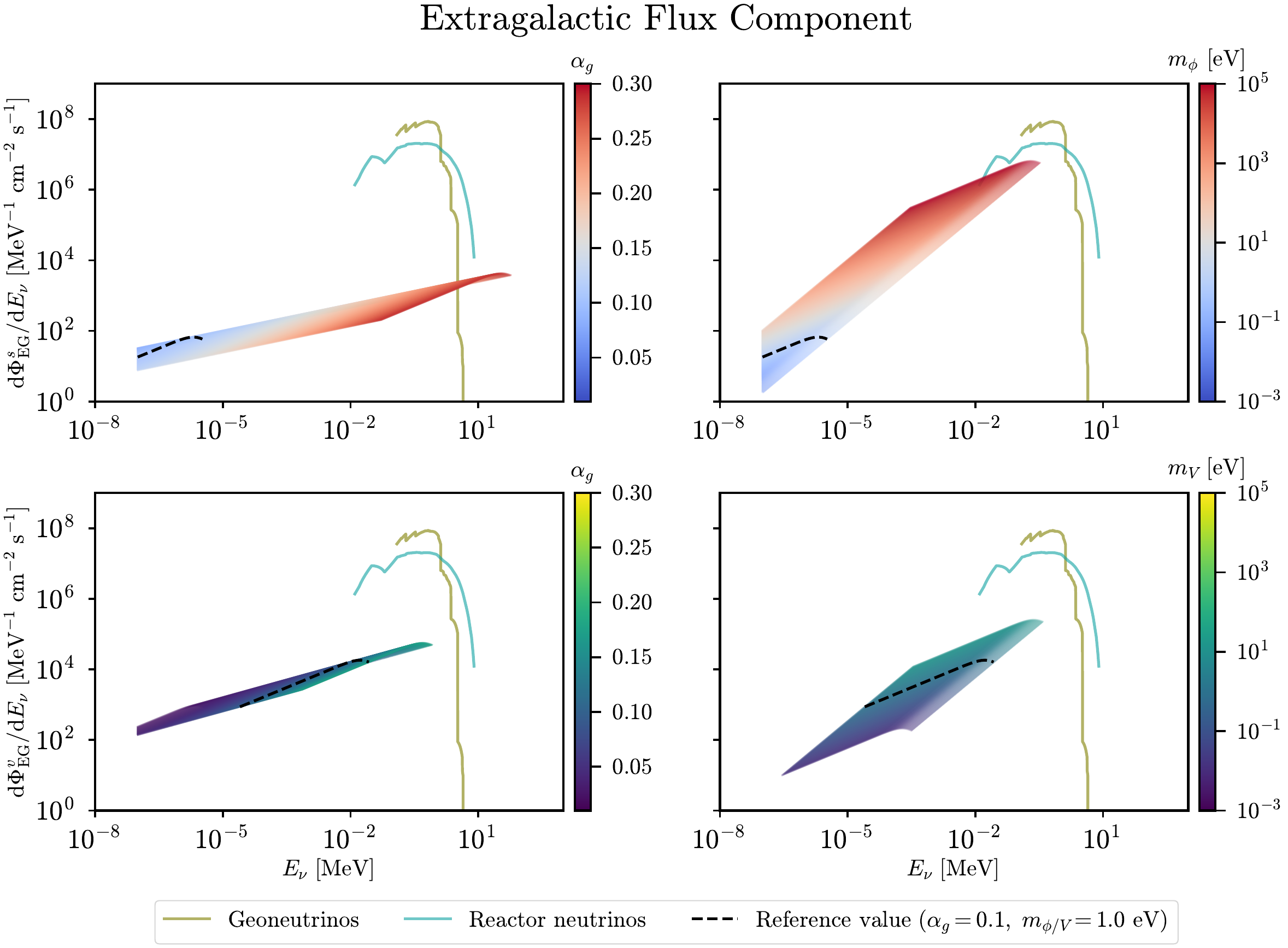}
    \caption{
    Example of the expected neutrino spectrum on Earth
    from PBH DM with a surrounding boson cloud. In the left column, 
    we fix the boson mass to the reference value and vary $\ag$
    while in the right column we fix $\ag$ on the reference value
    and vary the boson mass. All other values are set on the reference
    values; in particular, we set $\fpbh = 1$ here.
    For comparison with known fluxes
    we show the geoneutrino and the reactor flux
    from \cite{Vitagliano:2019yzm}.
    }
    \label{fig:EG_flux_varying}
\end{figure}

In \cref{fig:EG_flux_varying}, we show the expected differential neutrino flux
for our reference parameter choices, varying the boson mass and $\ag$ independently.
For comparison, we also include several known antineutrino fluxes at Earth, using the
results of Ref.~\cite{Vitagliano:2019yzm}. We focus on antineutrinos because, in our setup,
neutrinos and antineutrinos are produced in pairs, leading to equal fluxes for the two
components. In the relevant energy window, however, solar neutrinos constitute a large
foreground, making the antineutrino channel a cleaner target. From these plots, we can
already anticipate that the scalar case can lead to non-trivial constraints with existing
data, while the vector case is largely hidden below other flux components.

\subsection{Galactic Flux}
\label{subsec:galflux}

We now turn our attention to the Galactic flux. For this, we ignore the redshift dependence since it
is only a minor  correction. Instead, the spatial distribution of dark matter in the Milky Way must
be taken into account. In particular, the dark-matter density is expected to be larger toward the
Galactic center than in the outer halo. To model this distribution,  we will just assume here the standard
Navarro--Frenk--White (NFW) profile~\cite{Navarro:1996gj}
\begin{equation}
    \label{eqn:rho_NFW}
    \rho_{\text{NFW}}(r) = \frac{\rho_s}{\frac{r}{r_s}\left(1 + \frac{r}{r_s}\right)^2} \;,
\end{equation}
where $r_s = 24.42$~kpc is the scale radius. The normalization
$\rho_s$ is determined from the local density
\begin{equation}
    \rho_s = \rho_{\text{local}} \cdot x_0 (1+x_0)^2 \;, \quad x_0 = \frac{R_\odot}{r_s}
\end{equation}
with $R_\odot = 8.23$~kpc~\cite{R0value} the distance
of the solar system from the Galactic center and
for the local DM density we use
$\rho_{\text{local}} = 0.4$~GeV/cm$^3$~\cite{rholocal1_Catena:2009mf, rholocal2_Salucci:2010qr, rholocal3_Iocco:2015xga, rholocal4_Weber:2010LocalDM}. 

For a monochromatic population of PBHs of mass $\mpbh$ composing
a fraction $\fpbh$ of the Galactic DM,
the PBH number density at the galactocentric radius $r$ is given by 
\begin{equation}
    n_{\text{PBH}}(r) = \frac{\fpbh \, \rho_{\text{NFW}}(r)}{\mpbh} \;.
\label{eq:pbh_number_density}
\end{equation}
We  then calculate the neutrino flux from an angular direction
$\psi= (\ell , b)$ by calculating the line-of-sight integral
\begin{align}
    \frac{\diff \Phi_{\text{GA}}}{\diff E_\nu \diff \Omega}(E_\nu,\psi)
    &=
      \frac{1}{4\pi}\,
      Q_\nu(E_\nu) \,
      \int_0^\infty \diff s \,
      n_{\text{PBH}}(r(s,\psi))
      \nonumber \\[3pt]
    &=
      \frac{\fpbh}{4 \, \pi \, \mpbh}\,
      Q_\nu(E_\nu) \, J(\psi) \;,
    \label{eq:GalFluxJ}
\end{align}
with $Q_\nu$ from \cref{eq:Q_nu} and the standard $J$-factor
\begin{equation}
    J(\psi)
    = \int_0^\infty \diff s \,
      \rho_{\text{NFW}}(r(s,\psi)) \;.
\end{equation}
The galactocentric radius along the line-of-sight is given by
\begin{equation}
    \label{eq:r_los}
    r(s,\psi)
    =
    \sqrt{
        s^2 + R_\odot^2
        - 2 s R_\odot \cos b \cos \ell
    } \,.
\end{equation}

For the constraints considered below, we do not use directional information on the incoming neutrino flux.
We therefore work with the total Galactic flux, obtained by integrating over both neutrino energy and
solid angle, 
\begin{align}
    \label{eq:Phi_tot_GA}
    \Phi_{\text{GA}}^{\text{tot}} &= \int \frac{\diff \Phi_{\text{GA}}}{\diff E_\nu  \diff \Omega}(E_\nu,\psi) \diff E_\nu \diff \Omega
        = \frac{\fpbh}{4 \, \pi \, \mpbh}\,
      Q_0(E_\nu)  E_\nu \, J_\text{tot} \nonumber\\
      &= 3.52 \times 10^{33} \frac{\fpbh}{4 \, \pi \, \mpbh} \, Q_0(E_\nu) E_\nu \frac{\text{eV}}{\text{cm}^2} \;,
\end{align}
where we have defined the full-sky integrated  $J$-factor as 
\begin{align}
    J_\text{tot} = \int J \diff \Omega = 2.80 \times 10^{32} \frac{\text{eV}}{\text{cm}^2} \;.
\end{align}

In the expressions above, $Q_0$ is the same quantity as introduced in \cref{eq:Q_nu}. For the Galactic
contribution, redshift effects are negligible, so the observed neutrino energy is equal to the emitted
energy, $E_\nu=\bar{E}_\nu$. Therefore, in our setup, the Galactic flux appears as a line centered at
the emitted neutrino energy.

To estimate which component gives the larger contribution, Galactic or extragalactic, we compute the universal ratio
\begin{align}
    \Phi_{\text{GA}}^{\text{tot}} \left( \int \frac{\diff \Phi_{\text{EG}}}{\diff E_\nu \diff \Omega} \diff E_\nu \diff \Omega \right)^{-1} 
    &= \frac{J_\text{tot}}{16 \, \pi^2}\,
      \left( - \int \frac{g(r_\nu)}{r_\nu^2} \diff r_\nu 
      \right)^{-1} \approx 1.36 \;.
\end{align}

Within our setup, the dependence on the PBH and boson parameters cancels in this ratio. This comparison indicates
that the Galactic and extragalactic contributions are of similar magnitude, with the Galactic flux being
about $36\%$ larger. There is, however, a caveat. In this estimate, the extragalactic flux is integrated
over the full energy range. As shown in \cref{fig:EG_flux_varying}, only a fraction of the extragalactic
contribution lies within a given relevant energy interval. Therefore, in the energy bins where the Galactic
line contributes, the Galactic component is expected to be even more dominant.

\begin{figure}
    \centering
    \includegraphics[width=\linewidth]{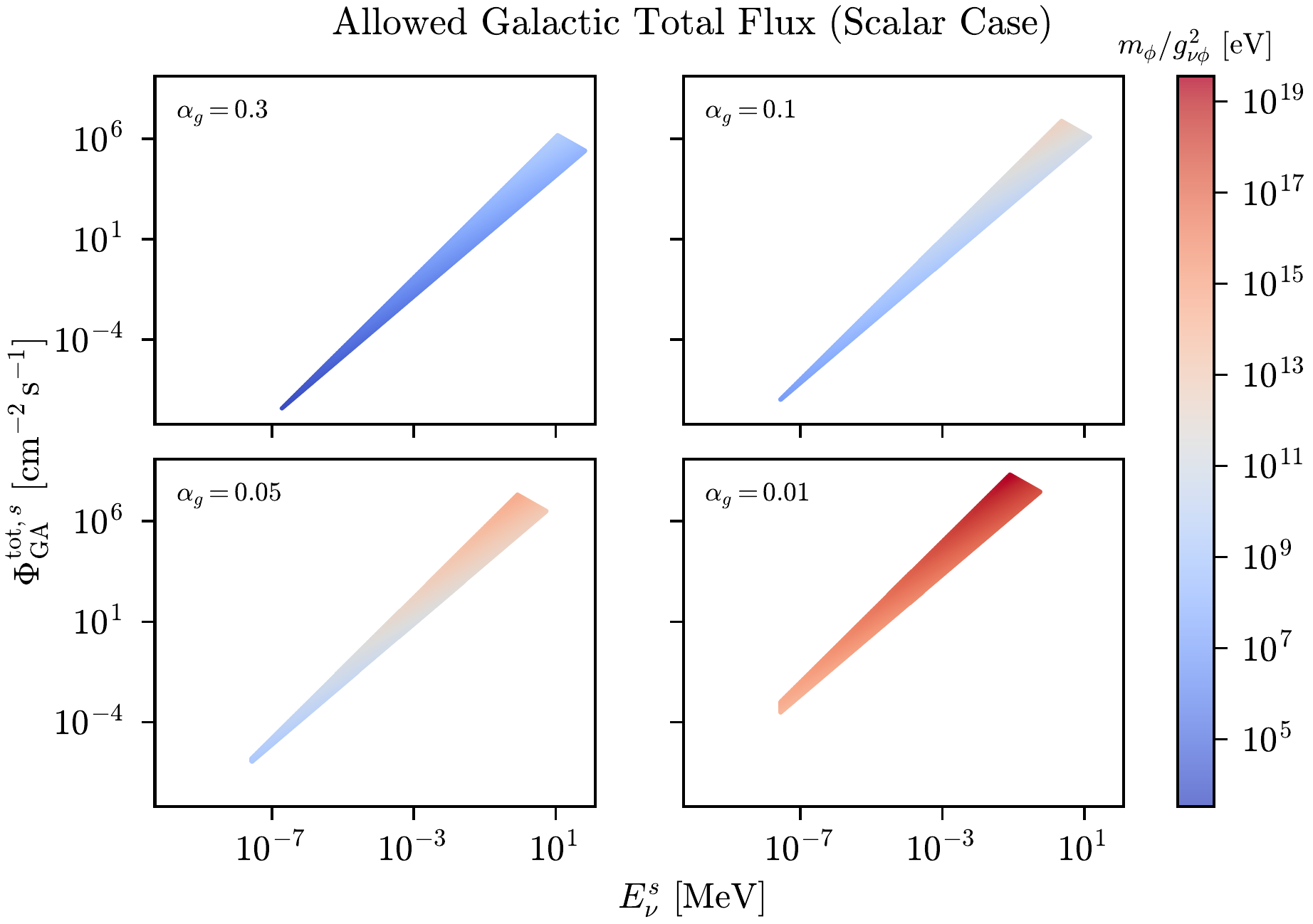}
    \caption{
        Allowed Galactic neutrino line flux $\Phi_{\text{GA}}^{\text{tot},s}$ on Earth as
        a function of the emitted neutrino energy $E_{\nu}^s$ for the scalar case. Each panel
        corresponds to a fixed value of $\ag$, while the color scale represents the combination
        $\mphi/\gvp^2$, with a common color normalization used across all panels. As discussed
        in the main text, the scalar Galactic flux satisfies 
        $\Phi_{\text{GA}}^{\text{tot},s}\propto \gvp^2 E_{\nu}^s$, which explains why
        the allowed points form a band. 
        }
    \label{fig:Galactic_scalar}
\end{figure}

\begin{figure}
    \centering
    \includegraphics[width=\linewidth]{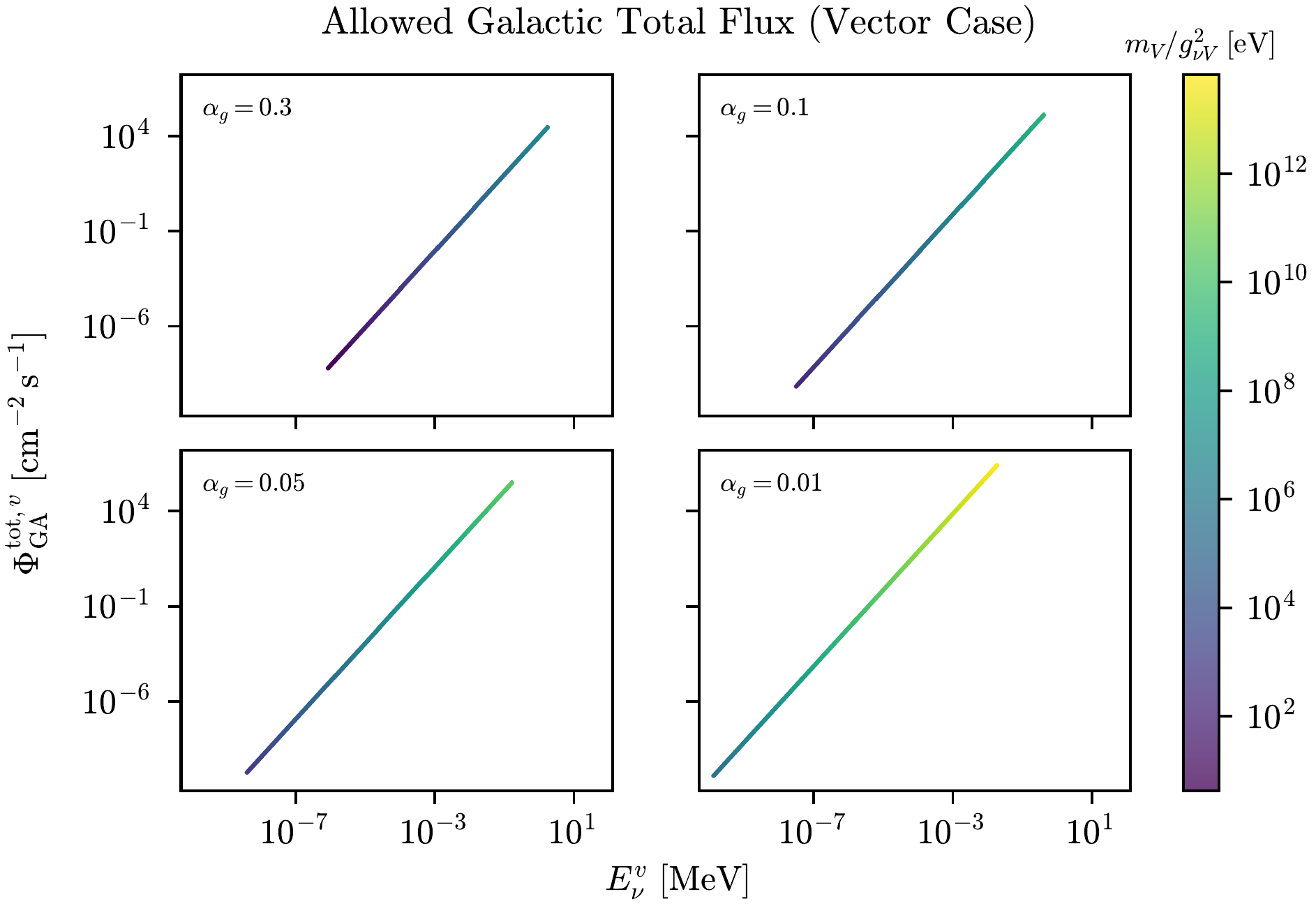}
    \caption{
        Allowed Galactic neutrino line flux $\Phi_{\text{GA}}^{\text{tot}, v}$ on Earth as
        a function of the emitted neutrino energy $E_{\nu}^v$ for the vector case. Each panel
        corresponds to a fixed value of $\ag$, while the color scale indicates the combination $\mV/\gvV^2$.
        Unlike for the single-PBH source parameter space, the allowed Galactic flux forms a narrow curve.
        This behavior follows from the fixed-$\ag$ scaling of the vector case, for which both $E_\nu^v$ and
        $\Phi_{\text{GA}}^{\text{tot},v}$ are controlled by the same parameter combination, leading to 
        $\Phi_{\text{GA}}^{\text{tot},v}\propto E_\nu^v$.
        }
    \label{fig:Galactic_vector}
\end{figure}

To give a sense of the size of the Galactic contribution, we show the total Galactic flux in
\cref{fig:Galactic_scalar} for the scalar case and in \cref{fig:Galactic_vector} for the vector case.
These figures are presented in the flux--energy plane, analogous to
\cref{fig:FluxEnergy_scalar_mphi_lines,fig:FluxEnergy_vector_mV_lines},
but now the flux corresponds
to the total flux observed at Earth. Unlike the source-level flux
regions shown earlier, the Galactic
contribution occupies a much narrower region in the scalar case 
and appears as a line
in the vector case. 

This behavior arises because the total Galactic flux contains the PBH number-density factor in
\cref{eq:pbh_number_density}, and therefore carries an additional dependence on $\mpbh$. Since $\ag$
is fixed in each panel, the relation between $\mpbh$ and the boson mass induces correlations among
the parameters. As a result, part of the parameter dependence becomes degenerate when projected onto
the flux--energy plane.
In the vector case, this degeneracy is particularly strong: both  the emitted neutrino energy and
the total Galactic flux are governed by the same parameter combination, so the allowed points 
collapse into a one-dimensional curve in the log-log plot.
We can make this a bit more quantitative. Using 
\cref{eq:Neutrino_Energy_Vector_no_PSi0,eq:Neutrino_Flux_From_One_Source_Vector_no_Psi0,eq:Q0,eq:Phi_tot_GA} 
we obtain
\begin{align}
    E_\nu^v \propto \frac{\mV}{\gvV^2} \text{ and } \Phi_{\text{GA}}^{\text{tot},v} \propto \frac{\mV}{\gvV^2}
\end{align}
(we fix all other parameters in the plots)
such that $\Phi_{\text{GA}}^{\text{tot}} \propto E_\nu^v$ 
which explains the linear behavior in \cref{fig:Galactic_vector}. This also motivates the choice
of $\mV/\gvV^2$ as the color scale.

Applying the same reasoning to the scalar case, we find
\begin{align}
    E_\nu^s \propto \frac{\mphi}{\gvp^4} \text{ and } \Phi_{\text{GA}}^{\text{tot},s} \propto \frac{\mphi}{\gvp^2} \;.
\end{align}

We therefore use $\mphi/\gvp^2$ as the color scale in the scalar case, since this combination controls
the total Galactic flux. Unlike in the vector case, however, the scalar flux and energy are not governed
by exactly the same parameter combination. Their ratio scales as 
$\Phi_{\text{GA}}^{\text{tot},s}/E_\nu^s \propto \gvp^2$, so for a fixed neutrino energy the flux
still depends on the remaining coupling $\gvp$.
As a result, the scalar points do not collapse onto a single line, but instead form a narrow band in
the flux--energy plane.

\section{Constraints on Parameter Space}
\label{sec:Constraints}

In the previous section, we discussed the theoretical expectations for the neutrino signal,
without yet comparing it in detail with experimental data. We found that the expected neutrino
energies can range from below the eV scale up to at most $\mathcal{O}(10\text{ MeV})$. Since
neutrinos with sub-MeV energies are extremely challenging to detect, we focus in the following
on the MeV energy range. In this regime, the dominant background is the large flux of solar neutrinos.
In principle, the signal considered here could be distinguished from solar neutrinos using directional
information, since in particular the Galactic contribution is expected to follow the dark-matter distribution rather
than the solar direction. In practice, however, such a separation may be challenging.

A more efficient strategy is to focus on antineutrinos. In this channel, the dominant backgrounds
come from geoneutrinos and reactor antineutrinos, whose fluxes are much smaller than the solar-neutrino
flux. Several experiments have placed limits on the antineutrino flux in the relevant energy range,
which can be recast as constraints on our signal. In particular, we use the results from
Borexino~\cite{Borexino:2010zht} in the range $2$--$18$~MeV,
KamLAND~\cite{KamLAND:2011bnd,KamLAND:2021gvi} in the range $8.3$--$18.3$~MeV,
and Super-Kamiokande \cite{Super-Kamiokande:2023xup} in the range $9.29$--$31.29$~MeV.
To compare our results with the experiments we first fix $\ag$ and $\gvp$.
Scanning over the boson mass then
corresponds to scanning over $\mpbh$ via \cref{eq:mphi_scaling}. For
each value of $\ag$, $\gvp$, and $\mpbh$ (we fix $\tilde{a} = 0.9$
and $N_\nu =1$), we  compute the corresponding differential 
extragalactic flux and the total Galactic neutrino flux.
Furthermore, for each energy bin constrained by the experiments listed above, we integrate the extragalactic flux over
this bin and add the Galactic line contribution whenever its energy falls inside the same bin.
We then impose the requirement that the predicted total flux does not exceed the corresponding
experimental $90\%$ C.L.\ upper limit in any relevant energy bin.

\begin{figure}
    \centering
    \includegraphics[width=\linewidth]{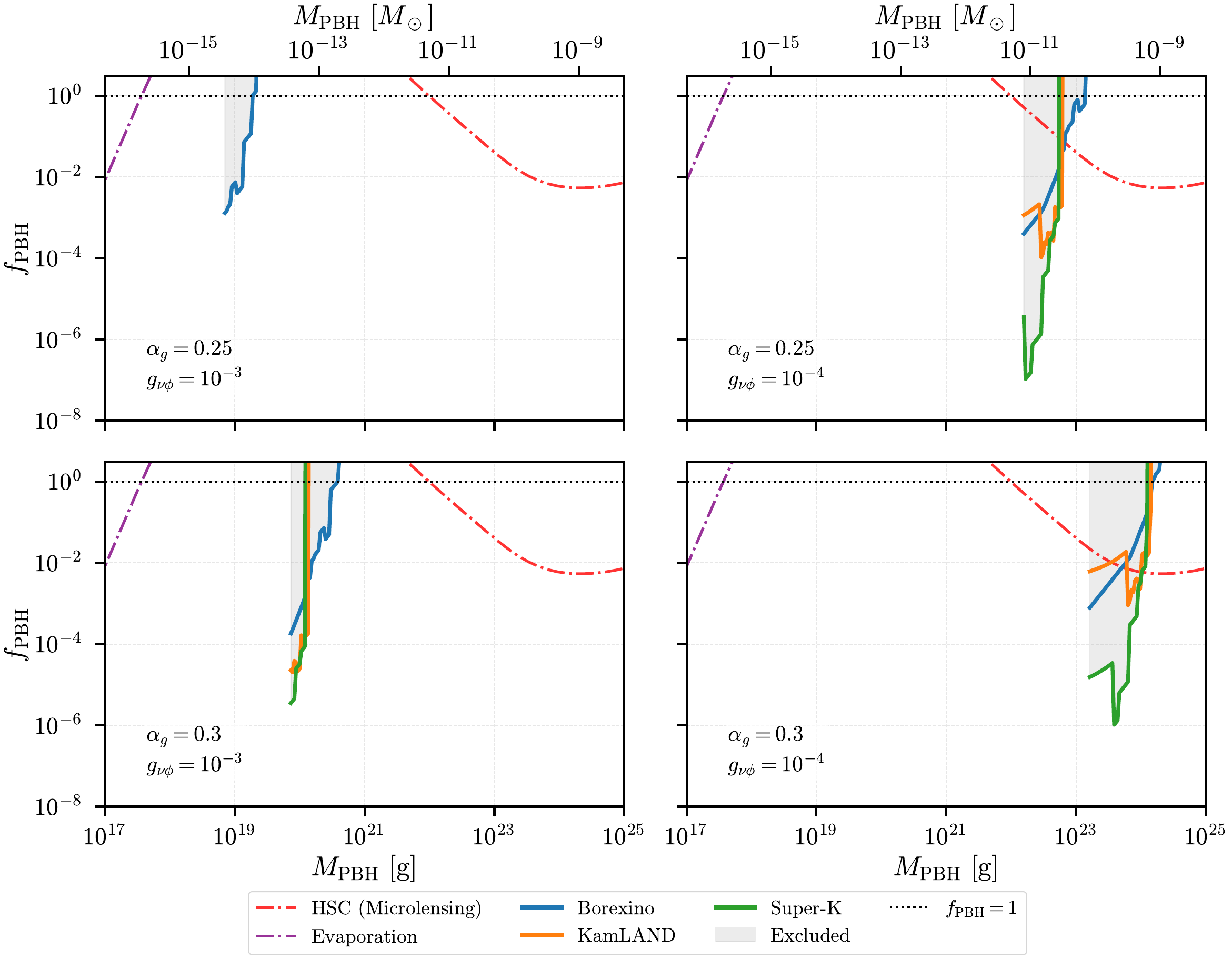}
    \caption{
    Upper limits on the PBH dark matter fraction $\fpbh$ for the 
    scalar case with $\alpha_g = 0.25$ (top), $\alpha_g = 0.3$ 
    (bottom), $\gvp = 10^{-3}$ (left), and
    $\gvp = 10^{-4}$ (right). We have set $\atilde = 0.9$
    and $N_\nu =1$ in all plots.
    The solid lines represent the constraints derived from the 
    neutrino superradiance mechanism presented here using 
    Borexino \cite{Borexino:2010zht} (blue), 
    KamLAND \cite{KamLAND:2011bnd, KamLAND:2021gvi} (red),
    and Super-Kamiokande \cite{Super-Kamiokande:2023xup} (green) 
    data. The shaded grey region is the combined bound.
    The pink dash-dotted line indicates the existing constraint
    from Subaru-HSC microlensing observations~\cite{Subaru-HSC:2020}
    and the purple dash-dotted line indicates Hawking evaporation
    constraints taken from \cite{bradley_j_kavanagh_2019_3538999}. 
    We remind here also about the relation \cref{eq:mphi_scaling} 
    which gives
    $\mphi \approx 2.7 \, \ag \, (10^{23} \text{ g}/ \mpbh)$~eV.
    }
    \label{fig:fpbh_vs_MPBH}
\end{figure}

We present our result in \cref{fig:fpbh_vs_MPBH} as $\fpbh$ vs.\ $\mpbh$ as it is commonly presented in the
PBH DM literature. We show constraints  for 
the benchmark values $\ag=0.25$, $0.3$ and $\gvp = 10^{-3}$, $10^{-4}$. For smaller values of $\ag$,
the constraint quickly disappears as the predicted neutrino energies fall below the energy ranges probed by
the experiments considered here. For larger values of $\ag$ we would need larger values of
$\gvp$, see \cref{fig:Parameters_Continuous_Flow}. But for larger values
of $\gvp$ the experimental constraints again become irrelevant which is apparent from comparing
our two benchmark values for this coupling. Furthermore,
the resulting bounds apply only to the scalar case. For the vector case,
the emitted neutrino energies are generally too low to yield competitive
constraints, as anticipated from \cref{fig:EG_flux_varying}.

The solid lines in \cref{fig:fpbh_vs_MPBH} represent the
constraints derived from our neutrino superradiance mechanism, using Borexino \cite{Borexino:2010zht} (blue), 
KamLAND \cite{KamLAND:2011bnd, KamLAND:2021gvi} (red)
and Super-Kamiokande \cite{Super-Kamiokande:2023xup} (green) data. The shaded grey region represents the
combined excluded region from these experiments.
The dash-dotted line indicates the existing constraint from  Subaru-HSC microlensing 
observations~\cite{Subaru-HSC:2020}.  This microlensing bound relies on the gravitational field
of the PBHs and not on the presence of any superradiance cloud. It is more model-independent in that regard.
Nevertheless, below about $\mpbh \sim 10^{24}$~g, our setup  provides significantly stronger bounds than the 
microlensing  limits.
Our main focus here is on the asteroid-mass window and \cref{fig:fpbh_vs_MPBH} shows
that this mechanism provides meaningful constraints for a significant portion of this range, from $\mpbh$ of a few times $10^{18}$~g all the way up to $10^{23}$~g.
The sensitivity even extends to somewhat larger masses, up to
$\mpbh \sim 10^{24}$~g. 
In particular, within the asteroid-mass range, our projected neutrino constraints
can surpass the corresponding Subaru-HSC microlensing bounds.

The PBH mass can be related to the scalar mass through \cref{eq:mphi_scaling}
which gives $\mphi \approx 2.7 \, \ag \, (10^{23} \text{ g}/ \mpbh)$~eV.
The excluded regions in \cref{fig:fpbh_vs_MPBH} then correspond to scalar masses
in the range from about $0.34 - 6.8$~eV for $\ag = 0.25$ and about $0.041 - 0.81$~eV 
for $\ag = 0.3$ respectively. We want to mention here that a larger fraction
of the PBH mass window might be excluded using other neutrino observables which
constrain their coupling to bosons around this mass window. However, such constraints often
depend on additional couplings of the boson to other particles, which would introduce model-dependent
assumptions beyond the scope of the present study. We therefore leave a detailed investigation of these
possibilities for future work.

\section{Summary and Conclusions} \label{sec:Summary}

Primordial Black Holes (PBHs) are a popular candidate for 
Dark Matter (DM) and in this paper we focus mostly on the asteroid
mass window, 10$^{17}$ -- 10$^{23}$~g,  
which has remained comparatively difficult
to constrain so far. For lighter masses, Hawking evaporation
provides significant constraints and for heavier masses
lensing constraints become relevant.  We have shown that, in the presence of light scalar bosons,
neutrino emission from superradiant clouds can lead to constraints that are significantly stronger
than existing lensing bounds for PBH masses $\mpbh \gtrsim 10^{22}$~g
and, in fact, constrain a wide range of the asteroid-mass window. We also considered
the case of vector bosons; however, in our setup the resulting neutrino energies and
fluxes are too small
to be constrained by current experiments.

The mechanism underlying these constraints relies on the presence of an additional light boson that can
form a superradiant cloud around rotating PBHs. If this boson couples to neutrinos, the cloud can emit
a steady neutrino flux through the mechanism discussed in Ref.~\cite{Chen:2023vkq}. We apply this idea
to asteroid-mass PBHs and compute the resulting Galactic and extragalactic neutrino fluxes at Earth.
We consider the region of parameter space in which the bosonic cloud can reach a saturated, or quenched,
state within the age of the Universe. In this phase, the energy extracted from the PBH through superradiance
is balanced by the energy carried away by neutrino emission, leading to an approximately steady neutrino
flux. Since the neutrino spectrum from a single PBH is sharply peaked, we model the source emission as a
monochromatic line, assuming universal couplings to the neutrino flavors.
We then discuss the propagation of the extragalactic and
Galactic neutrino flux to Earth. The extragalactic flux
has an energy distribution due to the redshift from the early universe, whereas the Galactic
contribution remains monochromatic. Within our setup, the ratio of the two contributions is independent
of the underlying PBH and boson parameters, with the Galactic flux expected to be about $36\%$ larger than
the extragalactic one.

At Earth, the predicted neutrino energies can span a wide range, from about $10^{-7}$~MeV up to nearly
$\mathcal{O}(100 \text{ MeV})$. The low-energy part of the spectrum is difficult to constrain because
of detector thresholds and large backgrounds from solar neutrinos, geoneutrinos, and reactor antineutrinos.
However, for part of the scalar parameter space, the predicted antineutrino flux exceeds existing bounds from
Borexino, KamLAND, and Super-Kamiokande. This allows us to derive the constraints
shown in \cref{fig:fpbh_vs_MPBH}.

For rapidly rotating PBHs with $\tilde{a}=0.9$ in the presence
of a light scalar particle, the strongest bound for $\alpha_g=0.25$ 
and $\gvp = 10^{-3}$ reaches roughly $\fpbh \sim 10^{-3}$ around
$\mpbh \sim 10^{19}$~g, while for $\alpha_g=0.3$ and $\gvp=10^{-3}$
it reaches $\fpbh \sim \text{few } \times 10^{-6}$ around 
$\mpbh \sim 10^{20}$~g. For the smaller Yukawa coupling 
$\gvp = 10^{-4}$, the strongest bound reaches 
$\fpbh \sim 10^{-7}$ around $\mpbh \sim 2 \times 10^{22}$~g 
for $\alpha_g=0.25$, and $\fpbh \sim 10^{-6}$ around 
$\mpbh \sim 4 \times 10^{23}$~g for $\alpha_g=0.3$. In the upper
part of this mass range, our neutrino bounds are significantly 
stronger than existing microlensing constraints. 

These results 
demonstrate that low-energy neutrino observations can provide a 
powerful and complementary probe of PBH DM in the asteroid-mass window.

\section*{Acknowledgments}

Yi-Xuan Lin and Martin Spinrath are supported by the National Science and Technology Council
(NSTC) of Taiwan under
Grant Nos.\ NSTC 114-2112-M-007-030 and NSTC 115-2112-M-007-028.

\bibliographystyle{JHEP}
\bibliography{paper_v1}

\end{document}